\documentclass[letter]{jpsj3}

\usepackage{bm}

\title{
Universal Properties of Nonlinear Response Functions\\
of Nonequilibrium Steady States
}

\author{
Akira \textsc{Shimizu}\thanks{E-mail address: shmz@ASone.c.u-tokyo.ac.jp}
}

\inst{
Department of Basic Science, 
University of Tokyo, 
Komaba, Meguro-City, Tokyo 153-8902, Japan
}

\abst{
We derive general properties, 
which hold in diverse physical systems, 
of nonlinear response functions of nonequilibrium steady states.    
In particular, sum rules and asymptotic behaviors are derived,
which relate many independent experiments.
Their consequences are illustrated for 
nonlinear optical materials and 
nonlinear electrical conductors.
}

\kword{nonequilibrium, nonlinear susceptibility, nonlinear response, sum rule, asymptotic behavior}

\begin{document}

\maketitle

The most important way of characterizing a physical state 
is to see its responses to weak external perturbations.
Such responses are generally represented by response functions. 
For equilibrium states, their linear ($\Phi^{(1) {\rm eq}}$)
and nonlinear (higher-order) 
response functions ($\Phi^{(2) {\rm eq}}, \Phi^{(3) {\rm eq}}, \cdots$)
have extensively been studied,
and many universal properties 
were discovered \cite{KTH,Zubarev,Shen}.
Regarding nonequilibrium states driven by a driving force $F$, 
in contrast,
only limited facts are known about universal properties of their 
response functions $\Phi^{(n)F}$ ($n=1, 2, \cdots$),
even for nonequilibrium steady states (NESSs) for which 
$\Phi^{(n)F}$'s are most clearly defined because
all macroscopic quantities take constant values 
(in the sense of eq.~(\ref{eq:steady}) below).

By universal properties we mean properties 
that are {\em measurable experimentally} (i.e., not formal)
and are {\em not limited to a specific class of systems}.
Unfortunately, neither the fluctuation theorem  \cite{ES2002}
(although it can reproduce known results 
for $\Phi^{(n) {\rm eq}}$ \cite{AG2007})
nor the Jarzynski equality \cite{JE}, 
both of which are considered to hold even in 
nonequilibrium, 
can give such universal properties of $\Phi^{(n)F}$
(at leat at present).
Furthermore, although many formal expressions
were previously derived for $\Phi^{(n)F}$ \cite{Ruelle},
such universal properties were not derived from them.

Universal properties of $\Phi^{(n) {\rm eq}}$ tend to be lost in 
$\Phi^{(n)F}$.
For example, the fluctuation-dissipation relations (FDRs)
and the reciprocal relations \cite{KTH,Zubarev}
are violated in NESSs \cite{YS2009,SY2010}.
Nevertheless, it was shown in ref.~\citen{SY2010} that 
several properties
hold universally for the {\em linear} response functions of NESSs, 
$\Phi^{(1)F}$, if those for $\Phi^{(1) {\rm eq}}$ are 
appropriately generalized. 
A natural question is: 
Do nonlinear $\Phi^{(n)F}$'s 
also have universal properties?

In this paper, we answer this question.
It is shown that 
the sum rules and the asymptotic behaviors hold universally 
if those for $\Phi^{(1)F}$  \cite{SY2010} are 
appropriately generalized to $n \geq 2$. 
The results are of fundamental importance to both physics 
and applications.

{\em NESS -- }
Suppose that a static driving force $F$, which can be strong, is applied to a 
{\em target system} (the macroscopic system of interest).
We consider the case where 
$F$ induces a NESS 
in the target system
for a sufficiently long time, i.e., for $[t_{\rm in}, t_{\rm out}]$,
where $t_{\rm out} - t_{\rm in}$ is macroscopically long. 
By a NESS we mean a state in which 
every {\em macroscopic} variable $A$ takes a constant 
value $\langle A \rangle_F$ 
in the sense that 
its expectation value at time $t$ behaves as \cite{SY2010}
\begin{equation}
\langle A \rangle_F^t = \langle A \rangle_F + 
o\left( \langle A \rangle_{\rm tp} \right).
\label{eq:steady}\end{equation}
Here, 
$\langle A \rangle_{\rm tp}$ denotes a typical value of $A$,
and 
$o\left( \langle A \rangle_{\rm tp} \right)$ represents
a (time-dependent) term which is negligibly small in the sense that 
$o\left( \langle A \rangle_{\rm tp} \right) / \langle A \rangle_{\rm tp}
\to 0$ as $V \to \infty$,
where $V$ denotes the volume of the target system.
When $A$ is the total spin $\vec{S}$, for example, 
$\langle \vec{S} \rangle_{\rm tp} = O(V)$ and
$\langle \vec{S} \rangle_F^t = \langle \vec{S} \rangle_F + o\left( V \right)$.
According to experiences, such NESSs appear in diverse physical systems.

We assume that {\em NESSs are stable against small perturbations}
because otherwise reproducibility of experiments would be lost.
That is, after small perturbations are removed
the target system returns to the same NESS 
as that before they were applied.
NESSs of most systems, 
apart from few exceptions such as glass near a melting point,
satisfy this assumption.

{\em Response functions of NESS -- }
Suppose that weak and time-dependent {\em probe fields} 
$f_1(t), f_2(t), \cdots, f_m(t)$ ($\equiv \boldsymbol{f}(t)$)
are applied, {\em in addition to $F$}, 
to the target system for $t \geq t_0$.
Here, $t_{\rm in} < t_0 \leq t < t_{\rm out}$.
The number $m$ of the probe fields is arbitrary,
although the case of $m=n$ is sufficient
for studying $\Phi^{(n)F}$ most generally.

We are interested in the response of the NESS to $\boldsymbol{f}(t)$.
Specifically, we focus on the response,
\begin{equation}
\Delta A(t)
\equiv 
\langle A \rangle_{F+\boldsymbol{f}}^t - \langle A \rangle_F,
\end{equation}
of a {\em macroscopic} dynamical variable $A$ 
of the target system.
Since we are considering a stable NESS, 
$\Delta A(t)$ may be expanded in powers of $\boldsymbol{f}$ 
({\em not} of $F$)
as $\Delta A(t) = A^{(1)}(t) + A^{(2)}(t) + \cdots$.
The $n$-th order response can then be expressed phenomenologically 
as \cite{Shen,ON1988,note:NLresponse}
\begin{equation}
\Delta A^{(n)}(t)
=
{1 \over n!} \sum_{\alpha_1=1}^m \cdots \sum_{\alpha_n=1}^m
\int_{t_0}^t d t_1 \cdots \int_{t_0}^t d t_n 
\Phi^{(n)F}_{\alpha_1 \cdots \alpha_n}(t-t_1, \cdots, t-t_n) 
f_{\alpha_1}(t_1) \cdots f_{\alpha_n}(t_n).
\label{eq:nth.response}
\end{equation}
As in the case of $\Phi^{(n){\rm eq}}$,\cite{Shen} 
the {\em $n$-th order response function} $\Phi^{(n)F}$ 
of the NESS is defined by 
this and the causality relation,
\begin{equation}
\Phi^{(n)F}_{\alpha_1 \cdots \alpha_n}(\tau_1, \cdots, \tau_n) 
= 0 \mbox{ if either one of } \tau_j\mbox{'s} <0, 
\label{eq:causality}\end{equation}
and by the requirement (to remove arbitrariness) that 
$
\Phi^{(n)F}_{\alpha_1 \cdots \alpha_n}(\tau_1, \cdots, \tau_n)
$
is {\em invariant under every permutation of 
$\alpha_1 \tau_1, \cdots, \alpha_n \tau_n$}.
As $F \to 0$, $\Phi^{(n)F}$ reduces to 
the $n$-th order response function $\Phi^{(n) {\rm eq}}$ 
of {\em equilibrium states} \cite{KTH,Zubarev,Shen}.

{\em Response to sinusoidal fields -- }
In many experiments \cite{Shen,ON1988},
the probe fields are taken as sinusoidal ones;
\begin{equation}
f_\alpha(t) 
=
f_\alpha^{+} e^{-i \omega_\alpha t}
+ f_\alpha^{-} e^{+ i \omega_\alpha t}
=
\sum_{\sigma = \pm 1} 
f_\alpha^{\sigma} e^{-i \sigma \omega_\alpha t},
\end{equation}
where $(f_\alpha^-)^* = f_\alpha^+$.
Then, eq.~(\ref{eq:nth.response}) gives 
the $n$-th order response as 
\begin{equation}
\Delta A^{(n)}(t)
=
{1 \over n!} \sum_{\alpha_1, \sigma_1} \cdots \sum_{\alpha_n, \sigma_n}
\Xi^{(n)F}_{\alpha_1 \cdots \alpha_n}
(\sigma_1 \omega_{\alpha_1}, \cdots, \sigma_n \omega_{\alpha_n})
f_{\alpha_1}^{\sigma_1} \cdots f_{\alpha_n}^{\sigma_n}
\ e^{ - i \left( \sigma_1 \omega_{\alpha_1} + \cdots 
+ \sigma_n \omega_{\alpha_n} \right) t}.
\end{equation}
Here, $\Xi^{(n)F}$ is the Fourier transform of 
$\Phi^{(n)F}$ \cite{range_of_tau};
\begin{equation}
\Xi^{(n)F}_{\alpha_1 \cdots \alpha_n}
(\sigma_1 \omega_{\alpha_1}, \cdots, 
\sigma_n \omega_{\alpha_n})
= 
\int_{- \infty}^{+\infty} \!\!\!\!\!\!\!\!\! d \tau_1
\cdots
\int_{- \infty}^{+\infty} \!\!\!\!\!\!\!\!\! d \tau_n
\
\Phi^{(n)F}_{\alpha_1 \cdots \alpha_n}(\tau_1, \cdots, \tau_n)
e^{ i \sum_{j=1}^n \sigma_j \omega_{\alpha_j} \tau_j }.
\label{def:Xi(n)}\end{equation}
When $n=m=2$, for example, 
$\Delta A^{(2)}(t)$ includes terms such as 
$(1/2)\Xi^{(2)F}_{1 1}(\omega_1, \omega_1) \left( f_1^+ \right)^2 
e^{ - 2 i \omega_1 t} + {\rm c.c.}$, 
$(1/2)\Xi^{(2)F}_{1 1}(\omega_1, -\omega_1) \left| f_1^+ \right|^2
 + {\rm c.c.}$,
$\Xi^{(2)F}_{1 2}(\omega_1, \omega_2) f_1^+ f_2^+
e^{ - i \left( \omega_1 + \omega_2 \right) t} + {\rm c.c.}$,
$\Xi^{(2)F}_{1 2}(\omega_1, -\omega_2) f_1^+ f_2^-
e^{ - i \left( \omega_1 - \omega_2 \right) t} + {\rm c.c.}$,
and so on.

{\em Microscopic expression of $\Phi^{(n)F}$ -- }
The equations so far presented are phenomenological ones
which are {\em closed in a macroscopic level}. 
We now relate them to {\em microscopic} physics by 
deriving a microscopic expression of $\Phi^{(n)F}$, 
which will be used to derive new universal properties.
In doing so, we do {\em not} employ perturbation expansion with respect to $F$
\cite{Zubarev,Shen}
because such expansion 
converges only slowly or does not converge for large 
$|F|$ of interest \cite{Zubarev,ON1988,SY2010},
except 
in limited physical situations.

\begin{figure}
\begin{center}
\includegraphics[width=.75\linewidth]{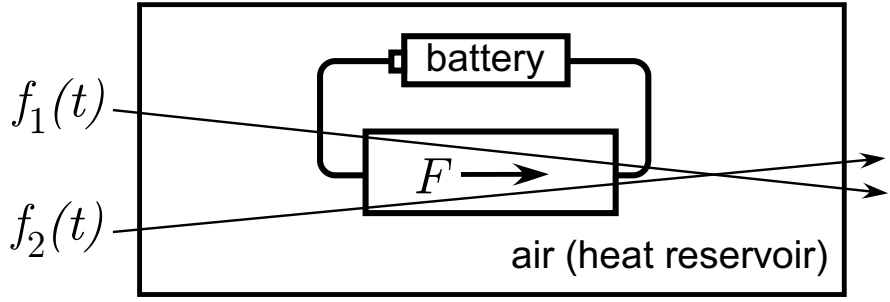}
\end{center}
\vspace{-3mm}
\caption{
An example of the `total system.'
It includes an electrical conductor (target system), 
a battery (source of the driving electric force $F$), 
air (heat reservoir), cables, and so on.
}
\label{fig:closed_NESS}
\end{figure}

To treat $F$ non-perturbatively,
we {\em tentatively} consider a huge system which includes 
the target system, a driving source that generates $F$, 
and their environments such as a heat reservoir(s), 
as shown in Fig.~\ref{fig:closed_NESS}.
We call this huge system the {\em total system}.
One can always take it large enough so that it
can be treated as a Hamiltonian system.
Its Hamiltonian is denoted by $\hat{H}^{\rm tot}$, 
which consists of 
the Hamiltonians of individual systems and interactions among them.
For $\hat{H}^{\rm tot}$, one should use not toy models
but a natural model which describes 
the real physical systems faithfully enough,
such as the full Hamiltonian (consisting of
electrons, nuclei and electromagnetic fields) for condensed matter.
Starting from such a huge system, 
we will finally derive relations which contain quantities 
{\em only of the target system}.

Regarding the probe fields $\boldsymbol{f}$ ($=f_1, f_2, \cdots$), 
we treat them as external weak fields,
which interact with the target system via the interaction term
$-\sum_{\alpha=1}^m \hat{B}_{\alpha} f_{\alpha}(t)$.
Here, $\hat{B}_{\alpha}$ is a macroscopic dynamical variable 
(which is the sum of local variables; see ref.~\citen{SM2005} and 
the later examples)
of the target system.
Hence, 
the density operator
of the total system $\hat{\rho}^{\rm tot}_{F+\boldsymbol{f}}(t)$ evolves 
as
\begin{equation}
i \hbar {\partial \over \partial t} 
\hat{\rho}^{\rm tot}_{F+\boldsymbol{f}}(t)
= 
\Big( 
\hat{H}^{\rm tot} - \sum_{\alpha=1}^m \hat{B}_{\alpha} f_{\alpha}(t)
\Big)^\times 
\hat{\rho}^{\rm tot}_{F+\boldsymbol{f}}(t).
\label{eq:vN_rho_tot}\end{equation}
Here, `$\times$' denotes the commutator; 
$\hat{a}^\times \hat{b} \equiv [\hat{a}, \hat{b}]$.

We denote $\hat{\rho}^{\rm tot}_{F+\boldsymbol{f}}(t)$
with $\boldsymbol{f}=0$ by $\hat{\rho}^{\rm tot}_{F}(t)$.
When $\boldsymbol{f}=0$, 
the reduced density operator of the target system is 
\begin{equation}
\hat{\rho}_F
\equiv
{\rm Tr}' \left[ \hat{\rho}^{\rm tot}_{F}(t) \right],
\label{eq:rhoF}\end{equation}
where ${\rm Tr}'$ denotes the trace operation over 
the degrees of freedom other than those of the target system.
As stated earlier, 
we consider the case where
a NESS is realized in the target system
(while the driving source is {\em not} in a steady state)
for a macroscopic time interval $[t_{\rm in}, t_{\rm out}]$.
In this interval, 
we can regard $\hat{\rho}_F$ as being independent of $t$
as far as macroscopic variables are concerned, 
\`a la eq.~(\ref{eq:steady}).
Unlike the equilibrium case, however, an explicit form of 
$\hat \rho_F$ is unknown.
Nevertheless, we can later derive 
universal relations among experimentally measurable quantities.

When $\boldsymbol{f} \neq 0$, 
$\hat{\rho}^{\rm tot}_{F}(t)$ 
is changed into 
$\hat{\rho}^{\rm tot}_{F+\boldsymbol{f}}(t)$.
For a macroscopic dynamical variable of interest, $\hat{A}$, 
of the target system, we see its response
$
\Delta A(t)
\equiv
\mbox{Tr} [ \hat{\rho}^{\rm tot}_{F+\boldsymbol{f}}(t) \hat{A} ] 
- 
\mbox{Tr} [ \hat{\rho}^{\rm tot}_{F}(t_0) \hat{A} ]
=
\mbox{Tr} [ \hat{\rho}^{\rm tot}_{F+\boldsymbol{f}}(t) \hat{A} ] 
- 
\mbox{Tr} [ \hat \rho_F \hat{A} ]
$.
Since we treat {\em stable} NESSs subject to {\em weak} $\boldsymbol{f}$,
we may evaluate $\Delta A(t)$  
by evaluating the solution of eq.~(\ref{eq:vN_rho_tot})
using a power-series expansion with respect to $\boldsymbol{f}$
(taking $\hat{\rho}^{\rm tot}_{F}(t)$ as the zeroth-order 
solution).
We then obtain the $n$-th order response as

\begin{eqnarray}
&&
\Delta A^{(n)}(t)
=
{1 \over n!}
\sum_{\alpha_1=1}^m \cdots \sum_{\alpha_n=1}^m
\int_{t_0}^t \! \! dt_1 
\cdots
\int_{t_0}^{t} \! \! dt_n 
{1 \over (i \hbar)^n}
\nonumber\\
&& \qquad \qquad \quad \times \
\mbox{Tr}
\left(
\hat{\rho}^{\rm tot}_{F}(t)
\vec{\mathcal{T}}
\left[
\breve{B}_{\alpha_n}(t_n-t)^\times
\cdots
\breve{B}_{\alpha_1}(t_1-t)^\times
\hat{A}
\right]
\right)
f_{\alpha_1}(t_1) \cdots f_{\alpha_n}(t_n),
\quad
\label{eq:DeltaA_soboku}
\end{eqnarray}
where $\vec{\mathcal{T}}$ is the chronological ordering operator,
and the symbol ` $\breve\null$ ' denotes an operator 
in the interaction picture, i.e., 
$
\breve B(\tau) \equiv 
e^{{i \over \hbar} \hat{H}^{\rm tot} \tau} \hat{B} \, 
e^{{-i \over \hbar} \hat{H}^{\rm tot} \tau}.
$
From consistency with the macroscopic physics, 
eq.~(\ref{eq:nth.response}), 
$t$ of $\hat{\rho}^{\rm tot}_{F}(t)$ in eq.~(\ref{eq:DeltaA_soboku})
must be irrelevant.
Hence, we can take it to be an arbitrary time (such as $t_0$)
in $[t_{\rm in}, t_{\rm out}]$,
and simply write $\hat{\rho}^{\rm tot}_{F}(t)$ as $\hat{\rho}^{\rm tot}_{F}$.
We thus obtain a general formula;\cite{note:NLresponse} 
\begin{equation}
\Phi^{(n)F}_{\alpha_1 \cdots \alpha_n}(\tau_1, \cdots, \tau_n)
=
{\mathcal{S}_{\bm{\alpha} \bm{\tau}} \over (i \hbar)^n}
\mbox{Tr}
\left(
\hat{\rho}^{\rm tot}_{F}
\left[
\breve{B}_{\alpha_n}(-\tau_n)^\times
\cdots
\breve{B}_{\alpha_1}(-\tau_1)^\times
\hat{A}
\right]
\right)
\label{eq:RCR}\end{equation}
for $\tau_1, \cdots, \tau_n \geq 0$.
Here, $\mathcal{S}_{\bm{\alpha} \bm{\tau}}$ denotes the operator
that symmetrizes the operand with respect to 
$\bm{\alpha} \bm{\tau}$ 
($\equiv \alpha_1 \tau_1,\, \cdots,\, \alpha_n \tau_n$),
i.e., 
$ 
\mathcal{S}_{\bm{\alpha} \bm{\tau}}
f(\alpha_1 \tau_1,\, \cdots,\, \alpha_n \tau_n)
=
(1/n!) 
\sum_{P} 
f(P(\alpha_1 \tau_1,\, \cdots,\, \alpha_n \tau_n)),
$
where $P$ denotes permutation.
Because of $\mathcal{S}_{\bm{\alpha} \bm{\tau}}$, 
$\vec{\mathcal{T}}$ in eq.~(\ref{eq:DeltaA_soboku})
becomes unnecessary in eq.~(\ref{eq:RCR}).

For classical systems, the commutators such as 
$
\breve{B}_{\alpha_1}(-\tau_1)^\times \hat{A}/i \hbar
$
should be replaced with the corresponding Poisson brackets.

The right-hand side (rhs) of eq.~(\ref{eq:RCR}) 
represents some correlation in the NESS.
However, 
even in the classical regime ($k_B T \gg \hbar \omega$),  
it is {\em not} equal to 
the time correlation of outcomes of experiments in general,
as explained for $n=1$ in ref.~\citen{SY2010}.
Hence, 
eq.~(\ref{eq:RCR}) is {\em not} the FDR.
We therefore call it simply 
the {\em $n$-th order response-correlation relation ($n$RCR)}.

{\em Simple relations -- }
Several universal relations can be derived 
{\em without} using the $n$RCR.
However, they are not so informative because 
they are nothing more than properties of Fourier transforms of 
real (causal) functions.
[Hence, they hold equally for $\Xi^{(n){\rm eq}}$ and $\Xi^{(n)F}$.]
For completeness, 
we present them before deriving new universal properties of $\Xi^{(n)F}$.

By definition, 
$
\Xi^{(n)F}_{\alpha_1 \cdots \alpha_n}
(\sigma_1 \omega_{\alpha_1}, \cdots, \sigma_n \omega_{\alpha_n})
$
is invariant under every permutation of 
$\alpha_1 \sigma_1 \omega_{\alpha_1}, \cdots, \alpha_n \sigma_n \omega_{\alpha_n}$,
where $\alpha_j = 1, 2, \cdots, m$
 ($j=1,2,\cdots,n$).
It also has the following obvious symmetry; 
\begin{equation}
\Xi^{(n)F}_{\alpha_1 \cdots \alpha_n}
(\sigma_1 \omega_{\alpha_1}, \cdots, \sigma_n \omega_{\alpha_n})
=
\Xi^{(n)F *}_{\alpha_1 \cdots \alpha_n}
(- \sigma_1 \omega_{\alpha_1}, \cdots, - \sigma_n \omega_{\alpha_n}).
\end{equation}
Hence, 
{\em its real and imaginary parts are even and odd functions, respectively}.
Furthermore, 
from eqs.~(\ref{eq:causality}) and (\ref{def:Xi(n)}), 
one can derive 
the {\em dispersion relations} \cite{Lucarini}. 
For $\omega_{\alpha_1}$, for example, 
\begin{equation}
\Xi^{(n)F}_{\alpha_1 \cdots \alpha_n}
(\sigma_1 \omega_{\alpha_1}, \sigma_2 \omega_{\alpha_2}, \cdots, \sigma_n \omega_{\alpha_n})
=
{-i \over \pi}
\int_{-\infty}^{\infty}
{
\mathcal{P}
\over
\omega' - \sigma_1 \omega_{\alpha_1}
}
\Xi^{(n)F}_{\alpha_1 \cdots \alpha_n}
(\omega', \sigma_2 \omega_{\alpha_2}, \cdots, \sigma_n \omega_{\alpha_n})
d \omega,'
\end{equation}
where $\mathcal{P}$ denotes the principal part.
By taking the real and imaginary parts of this relation,
one obtains relations between 
${\rm Re} \, \Xi^{(n)F}$ and ${\rm Im} \, \Xi^{(n)F}$.
One can also derive {\em moment sum rules} \cite{Lucarini} 
which are generalizations of 
those for $\Xi^{(1) {\rm eq}}$ \cite{KTH}.

{\it Sum rules and asymptotic behaviors of $\Xi^{(n) F}$ -- }
Using the $n$RCR, 
we now derive universal properties \cite{chigai} which
are very useful in physics and applications.
Their physical meanings, implications and typical examples 
will be discussed after presenting them all.

The first property is the following {\em sum rule for 
${\rm Re} \, \Xi^{(n)F}$};
\begin{equation}
\int_{-\infty}^{\infty} \!\! {d \omega_1 \over \pi} 
\cdots
\int_{-\infty}^{\infty} \!\! {d \omega_n \over \pi} 
\
{\rm Re} \, \Xi^{(n)F}_{\alpha_1 \cdots \alpha_n}
(\sigma_1 \omega_1, \cdots, \sigma_n \omega_n)
=
{\mathcal{S}_{\bm{\alpha}} \over (i \hbar)^n}
\left\langle  
\hat{B}_{\alpha_n}^{\ \times} \cdots \hat{B}_{\alpha_1}^{\ \times} \hat{A} 
\right\rangle _F, 
\label{sr:Re}\end{equation}
which {\em holds for all $F$} as far as a stable NESS is realized,
and, obviously, for all $\sigma_1, \cdots, \sigma_n$.
Here, 
$ 
\langle \cdot \rangle _F 
\equiv {\rm Tr} \left( \hat \rho_F \ \cdot \ \right) 
$ 
denotes the expectation value in the NESS of 
the target system, eq.~(\ref{eq:rhoF}).
To derive this sum rule, we note that 
the left-hand side (lhs) of eq.~(\ref{sr:Re}) equals 
the integral of $\Xi^{(n)F}$ because 
${\rm Im} \, \Xi^{(n)F}$ is an odd function.
Hence, the lhs equals  
$\Phi^{(n)F}_{\alpha_1 \cdots \alpha_n}(+0, \cdots, +0)$,
which 
is evaluated from eq.~(\ref{eq:RCR}) as the rhs
of eq.~(\ref{sr:Re}).

We also obtain the following {\em sum rule for 
${\rm Im} \, \Xi^{(n)F}$};
\begin{eqnarray}
&& 
\int_{-\infty}^{\infty} \!\! {d \omega_j \over \pi} 
\left[
\sigma_j \omega_j 
\left( \prod_{k \ (\neq j)} 
\int_{-\infty}^{\infty} \!\! {d \omega_k \over \pi} 
\right)
{\rm Im} \, \Xi^{(n)F}_{\alpha_1 \cdots \alpha_n}
(\sigma_1 \omega_1, \cdots, \sigma_n \omega_n)
-
{\mathcal{S}_{\bm{\alpha}} \over (i \hbar)^n}
\left\langle  
\hat{B}_{\alpha_n}^{\ \times} \cdots \hat{B}_{\alpha_1}^{\ \times} \hat{A} 
\right\rangle _F
\right]
\nonumber\\ 
&& \quad 
=
-
{\mathcal{S}_{\bm{\alpha}} \over (i \hbar)^n}
\left\langle  
\hat{B}_{\alpha_n}^{\ \times} \cdots 
\dot{\breve{B}}_{\alpha_j}(0)^\times \cdots
\hat{B}_{\alpha_1}^{\ \times} 
\hat{A} 
\right\rangle _F,
\qquad
\label{sr:Im}\end{eqnarray}
which {\em holds for all $F$} 
(as far as a stable NESS is realized)
and all $j$ ($=1,2,\cdots,n$),
and for all $\sigma_1, \cdots, \sigma_n$.
To obtain this relation for $j=1$ for example, 
integrate eq.~(\ref{def:Xi(n)}) by parts with respect to $\omega_1$, 
multiply the resulting equation with 
$\sigma_1 \omega_1$, 
integrate the resulting equation over $\omega_2, \cdots, \omega_n$, 
and finally use  eq.~(\ref{eq:RCR}).

From the above sum rules, 
we can also derive {\em asymptotic behaviors} of $\Xi^{(n) F}$ as follows.
Regarding ${\rm Re} \, \Xi^{(n)F}$, it should decay quickly for large 
$\omega_1, \omega_2, \cdots$
in such a way that 
the integral of eq.~(\ref{sr:Re}) converges
because the rhs of eq.~(\ref{sr:Re}) is finite.
Regarding ${\rm Im} \, \Xi^{(n)F}$, 
it should behave for large $\omega_j$ ($j=1,2,\cdots,n$) as
\begin{equation}
\left( \prod_{k \ (\neq j)} 
\int_{-\infty}^{\infty} \!\! {d \omega_k \over \pi} 
\right)
{\rm Im} \, \Xi^{(n)F}_{\alpha_1 \cdots \alpha_n}
(\sigma_1 \omega_1, \cdots, \sigma_n \omega_n)
\sim
{1 \over \sigma_j \omega_j}
{\mathcal{S}_{\bm{\alpha}} \over (i \hbar)^n }
\left\langle  
\hat{B}_{\alpha_n}^{\ \times} \cdots \hat{B}_{\alpha_1}^{\ \times} \hat{A} 
\right\rangle _F
\label{asym:Im}\end{equation}
{\em for all  $F$} 
(as far as a stable NESS is realized)
and all $j$ ($=1,2,\cdots,n$),
and for all $\sigma_1, \cdots, \sigma_n$,
because the rhs of eq.~(\ref{sr:Im}) is finite.
When the multiple commutator in the rhs of eq.~(\ref{asym:Im}) vanishes, 
${\rm Im} \, \Xi^{(n)F}$ decays more quickly.

{\it Physical meanings and implications -- }
As discussed in ref.~\citen{SY2010} for $n=1$, 
the above sum rules should be considered as predictions 
not on $\Phi^{(n)F}$ at $\tau_j \to +0$
but on $\Xi^{(n)F}$ at many different frequencies 
(in a certain finite range, as discussed in ref.~\citen{SY2010}),
because experiments are hard for the former and easy for the latter.

The above sum rules and asymptotic behaviors contain
quantities {\em only of the target system} (and its boundaries), 
namely, 
$\hat{A}, \hat{B}_{\alpha_1}, \cdots,  
\hat{B}_{\alpha_n}, \hat{\rho}_F
$
and $\dot{\breve{B}}_{\alpha_j}(0)$ \cite{note:dotB}.
Since all of these operators except for 
$\hat{\rho}_F$ are known operators,
the nonequilibrium averages in these relations 
can easily be measured experimentally.
For example, to measure 
$
\left\langle  
\hat{B}_{\alpha_n}^{\ \times} \cdots \hat{B}_{\alpha_1}^{\ \times} \hat{A} 
\right\rangle _F
$,
all one has to do is simply to measure 
$\hat{B}_{\alpha_n}^{\ \times} \cdots \hat{B}_{\alpha_1}^{\ \times} \hat{A}$
(which often reduces to a simple operator, as will be 
illustrated later)
in the NESS {\em without applying $\boldsymbol{f}(t)$}.
One can also measure $\Xi^{(n)F}$ easily 
by measuring the responses {\em by applying $\boldsymbol{f}(t)$}
of various frequencies.
Therefore, 
{\em all terms in our universal relations can easily 
be measured experimentally}.
Furthermore, 
the relations give {\em predictions on two or more independent experiments}
(e.g., one with $\boldsymbol{f}(t)$ and another without $\boldsymbol{f}(t)$).

Let us examine a typical case where 
$\hat{A}$ is an $l_A$-th order polynomial of momentum (or position) variables
whereas $\hat{B}_{\alpha_j}$'s are 
$l_B$-th order polynomials of position (or momentum) variables.
Then,
$\hat{B}_{\alpha_n}^{\ \times} \cdots \hat{B}_{\alpha_1}^{\ \times} \hat{A}$
is an $[(l_B-2)n + l_A]$-th order polynomial.
When $l_B = 1$, in particular, 
it becomes a lower-order polynomial for higher $n$, until it vanishes 
for $n > l_A$.
Hence, we are led to a remarkable conclusion that 
{\em if $l_B = 1$ the sum value of eq.~(\ref{sr:Re}) 
and the asymptotic value of eq.~(\ref{asym:Im})
become completely independent of $F$ for $n \geq l_A$}.

For nonlinear optical susceptibilities, for example, 
$l_A=l_B=1$ because $\hat{A}$ and $\hat{B}$ are proportional to 
the sums of positions and momenta of electrons, respectively.
Hence, 
the sum and asymptotic values are independent of 
$F$ for all $n$ ($\geq 1$).

Note that 
we have made almost no assumption except that the NESS is stable.
Although 
the reduced (projected) dynamics of the {\em target} system 
may be described by 
non-Hamiltonian models such as dissipative stochastic models, 
one can always take the {\em total} system 
(such as Fig.~\ref{fig:closed_NESS})
large enough so that it is well described as a Hamiltonian system.
Our universal results
have been derived from 
the Hamiltonian dynamics of such a huge system.
[Nevertheless, they are relations among quantities 
only of the target system.]
Therefore, 
{\em our results should hold in diverse physical systems},
including electrical conductors, optical 
materials, magnetic substances, organic materials, and so on,
{\em even when they are subject to dissipative environments}.

{\em Nonlinear electrical conductors --}
As an illustration, 
consider the case where the target system is 
a nonlinear electrical conductor of length $L$.
Let $e$ and $m$ be electron's charge and mass, respectively, 
and $\hat{q}_x^j$ and $\hat{p}_x^j$ be the $x$ components of the position and
momentum, respectively, of the $j$th electron in the conductor.
We here consider the simplest case where 
both the static electric field $F/e$ and the 
probe electric fields $\boldsymbol{f}(t)$
are applied along the conductor, in the $x$ direction.
Hence, $\hat{B}_\alpha = \sum_{j} e \, \hat{q}_x^j$ 
($\equiv \hat{B}$) for all $f_\alpha$'s,
and $l_B=1$.

If one is interested in 
the electric current averaged over the $x$ direction, 
whose operator may be 
$
\hat{I} \equiv (e/mL)\sum_{j} \hat{p}_x^j
$,
then putting
$\hat{A} = \hat{I}$ yields 
$\hat{B}^\times \hat{A}/(i \hbar) = (e^2 N_e/mL) \hat{1}$.
Here, $N_e$ is the number of electrons in the conductor,
and $\hat{1}$ is the identity operator.
Hence, $\hat{B}^\times \hat{B}^\times \hat{A}=0$.
These are consistent with the above result on a typical case,
where $l_A = l_B =1$ in this example.
The sum rule, eq.~(\ref{sr:Re}), 
for the $n$-th order response function of 
$
\Delta I(t)
=
\langle I \rangle_{F+\boldsymbol{f}}^t - \langle I \rangle_F
$
reads
\begin{equation}
\mbox{integral in eq.~(\ref{sr:Re})}
=
\begin{cases}
e^2 N_e/m L
& (n=1),\\
0
& (n \geq 2).
\end{cases}
\label{sum.for.I}\end{equation}
Hence, {\em the sum value is independent of $F$}. 
This is remarkable because  
${\rm Re} \, \Xi^{(n) F}$ at individual values of $\omega_\alpha$'s 
depends strongly on $F$ at low frequencies \cite{SY2010}
(because, e.g., electrons become hot by $F$).
The sum rule implies that 
such strong dependence on $F$ at low frequencies is {\em always} 
canceled completely,
after integration over frequencies, by 
weak dependence at high frequencies.
Regarding the asymptotic value of eq.~(\ref{asym:Im}),
it equals the above sum value divided by $\sigma_j \omega_j$.

On the other hand, 
if one is interested in the kinetic energy $\hat{K}$
(to evaluate, e.g., the  `kinetic temperature'),
then putting
$\hat{A} = \hat{K}
=
(1/2m) 
\sum_j \left[ (\hat{p}_x^j)^2 + (\hat{p}_y^j)^2 + (\hat{p}_z^j)^2 \right]
$ 
yields
$\hat{B}^\times \hat{A}/(i \hbar) = (e/m) \sum_j \hat{p}_x^j = L \hat{I}$,
$\hat{B}^\times \hat{B}^\times \hat{A}/(i \hbar)^2 
= (e^2 N_e / m) \hat{1}$,
and
$\hat{B}^\times \hat{B}^\times \hat{B}^\times \hat{A} = 0$.
These are consistent with the above result on a typical case,
where $l_A =2, l_B =1$ in this example.
The sum rule, eq.~(\ref{sr:Re}), 
for the $n$-th order response function of 
$
\Delta K(t)
=
\langle K \rangle_{F+\boldsymbol{f}}^t 
- \langle K \rangle_F
$
reads
\begin{equation}
\mbox{integral in eq.~(\ref{sr:Re})}
=
\begin{cases}
L \langle I \rangle_F
& (n=1), \\
e^2 N_e / m
& (n=2), \\
0
& (n \geq 3).
\end{cases}
\label{sum.for.K}\end{equation}
In this case, the sum value depends strongly on $F$ for 
$n=1$, 
whereas it is independent of $F$ for $n \geq 2$.
Regarding the asymptotic value of eq.~(\ref{asym:Im}),
it equals the above sum value divided by $\sigma_j \omega_j$.

These results demonstrate that 
{\em whether the sum value depends on $F$ is determined 
by the observable of interest $\hat{A}$, 
the operators $\hat{B}_\alpha$'s which couple to $\boldsymbol{f}$,
and the order $n$ of the response}.
We can say the same for the asymptotic values given by eq.~(\ref{asym:Im}).

{\em Concluding remarks -- }
As discussed above, the present results
hold in diverse physical systems, both quantum and classical, 
even when they are subject to dissipative environments.
Hence, they 
will become foundations of nonlinear statistical mechanics
and condensed-matter physics of NESSs, 
in the same way as 
the universal properties of $\Phi^{(1) {\rm eq}}$ are foundations 
in the linear nonequilibrium regime
\cite{KTH,Zubarev}.

The present results
are also important to applications
because nonlinear responses are widely used in 
electrical and optical engineering \cite{ON1988,FS1992}.
Since NESSs contain equilibrium states as the limiting case $F \to 0$,
$\Phi^{(n)F}$ has greater potential than  $\Phi^{(n) {\rm eq}}$.
Our results show that there exist fundamental limits 
to the spectra of $\Phi^{(n)F}$ 
even if one uses NESSs instead of equilibrium 
states.
For example, as shown above, 
the integral of nonlinear optical susceptibilities 
over frequencies does not change by application of $F$,
however large $F$ is.

Experimentally, 
the present results can be verified  by 
measuring $\Xi^{(n)F}$ over a wide frequency range.
One can also confirm the present results by molecular dynamics 
simulations, as was done for $n=1$ in ref.~\citen{SY2010}.

Conversely, one can use the present results 
to examine the correctness of results of experiments or 
theoretical calculations,
in the same way as one uses the charge conservation to 
examine the correctness of his results.

The author thanks T. Yuge for helpful discussions.
This work was supported by KAKENHI No.~22540407.


\begin{thebibliography}{99}
\bibitem{range_of_tau}
One can make the upper limit ($=t_{\rm out} - t_{\rm in}$) of $\tau_j$'s 
arbitrarily large by 
increasing the size of the source of $F$.
\bibitem{chigai}
Relations which might look similar to eqs.~(\ref{sr:Re}) and (\ref{asym:Im})
were derived in ref.~\citen{Lucarini}. 
However, 
$\omega_1=\omega_2=\cdots$ was assumed there.
More importantly, 
since they are general properties of Fourier transforms, 
no information was given on the sum and asymptotic values.
That is, their dependence on $F$, states, or system parameters 
was not given, although this is the central question in 
nonequilibrium physics.
In contrast, our results give the complete answer to this central question.
\bibitem{note:dotB}
$\dot{\breve{B}}_{\alpha_j}(0)$
($= [\hat{B}_{\alpha_j}, \hat{H}^{\rm tot}]/i \hbar$) 
includes operators only 
of the target system and its boundaries
because, physically, all (elementary) interactions in  $\hat{H}^{\rm tot}$
should be local interactions.
\end{thebibliography}
\end{document}